\documentclass[prb,aps,10pt,twocolumn,superscriptaddress,showpacs,preprintnumbers,amsmath,amssymb]{revtex4-1}

\usepackage{graphicx}
\usepackage{dcolumn}
\usepackage{bm}
\usepackage{here}

\usepackage{ulem}

\begin{document}
\title{Reentrant topological transitions in a quantum wire/superconductor system with quasiperiodic lattice modulation}
\author{Masaki Tezuka}
\email{tezuka@scphys.kyoto-u.ac.jp}
\affiliation{Department of Physics, Kyoto University, Kitashirakawa, Sakyo-ku, Kyoto 606-8502, Japan}
\author{Norio Kawakami}
\affiliation{Department of Physics, Kyoto University, Kitashirakawa, Sakyo-ku, Kyoto 606-8502, Japan}
\date{\today}

\begin{abstract}
We study the condition for a topological superconductor (TS) phase with end Majorana fermions
to appear when a quasiperiodic lattice modulation is applied to
a one-dimensional quantum wire with strong spin-orbit interaction
situated under a magnetic field and in proximity to a superconductor.
By density-matrix renormalization group analysis,
we find that multiple topological phases with Majorana end modes
are realized in finite ranges of the filling factor,
showing a sequence of reentrant transitions as the chemical potential is tuned.
The locations of these phases reflect the structure of bands in the non-interacting case,
which exhibits a distinct self-similar structure.
The stability of the TS in the presence of
an on-site interaction or
a harmonic trap potential is also discussed.
\end{abstract}

\maketitle

Edge states of topologically nontrivial
systems have attracted attention
because they are \textit{topologically protected}, that is, they are stable against weak
perturbations that do not change the topology of their quantum state.
Majorana surface states can form at the boundaries or vortex cores of topological superconductors (TSs),
\cite{Read2000, Kitaev2001, Ivanov2001, Fu2008, Sato2009, Linder2010}
and there has been a large amount of effort to observe such states
partly because they are expected to be useful in realizing quantum computation. \cite{Kitaev2001, Tewari2007}
The TS can form in a one-dimensional (1D) quantum wire with
spin-orbit interaction (SOI) that is placed under a Zeeman field and close to a bulk superconductor, \cite{Sau2010, Alicea2010, Lutchyn2010, Oreg2010}
or in cold atoms in two-dimensional optical lattices with
effective gauge fields generated by spatially varying laser fields, \cite{Sato2009} among others.

Here we are interested in the effect of spatial inhomogeneity of the system, \cite{Motrunich2001,Gruzberg2005}
imposed as a modification of site energy levels,
on the realization of Majorana end states.
Recently Brouwer and coworkers \cite{Brouwer2011} have studied two models,
the Dirac equation with random mass and a 1D spinless
superconductor, and obtained the energy distributions of the end states.
The end modes of non-interacting 1D systems with quasiperiodic
potentials have also been studied in connection with two-dimensional integer
quantum Hall systems \cite{Lang2011} and quantum spin Hall systems. \cite{Mei2012}

There is also a growing interest in the role of the electron-electron interaction in
1D conductors with SOI and topological materials. \cite{Braunecker2010, Fidkowski2010, Gangadharaiah2011, Stoudenmire2011, Sela2011, Lutchyn2011b}
Stoudenmire and coworkers \cite{Stoudenmire2011} showed that
while an on-site electron-electron interaction $U$
reduces the proximity-induced superconducting gap on a quantum wire attached to a
bulk superconductor, the chemical potential range of TS
with Majorana end states is enlarged by $U>0$.

In this work we study the effect of a quasiperiodic site level modification
of a 1D lattice. Such a system has been realized in cold-atom
systems \cite{Roati2008, Lucioni2011} and can also be relevant when the quantum wire is
placed on a bulk superconductor which has the lattice (or superlattice)
constant incommensurate with that of the quantum wire.
We find that TS with Majorana end states are observed in several regions with
finite widths of the chemical potential closer to the band center,
which resembles the multichannel case. \cite{Lutchyn2011b, Potter2010, Lutchyn2011a}
Moreover those TS regions are broadened by $U>0$.
This result paves a way to the observation and manipulation of Majorana end states
in various 1D TS systems with inhomogeneities, which may be tunable or unavoidable.

\noindent\textit{Setup.---}
We study a tight-binding 1D fermion model by the 
density-matrix renormalization group (DMRG). \cite{White1992, Schollwock2011}
Up to 160 eigenstates of the reduced density matrix $\rho$,
with the sum of discarded eigenvalues of $\rho$ at each step in the last finite-size
system loop being typically less than $10^{-7}$, have been retained in the DMRG calculation.

We adopt the following Hamiltonian,
\begin{eqnarray}
\mathcal{H}
&=& -\frac{t}{2}\sum_{l=0}^{L-2}\sum_{\sigma=\uparrow,\downarrow}
(\hat c_{\sigma,l}^\dag \hat c_{\sigma,l+1} + \mathrm{h.c.})\nonumber\\
&+& U \sum_{l=0}^{L-1} \hat n_{\uparrow,l} \hat n_{\downarrow,l}
+ \Delta \sum_{l=0}^{L-1}(\hat c_{\uparrow,l} \hat c_{\downarrow,l} + \mathrm{h.c.})\nonumber\\
&+& \sum_{l=0}^{L-1} \left(V_z(\hat n_{\uparrow,l}-\hat n_{\downarrow,l})
+\sum_{\sigma=\uparrow,\downarrow}(t-\mu+\epsilon_{\sigma,l}) \hat n_{\sigma,l}\right)\nonumber\\
&+& \frac{\alpha}{2} \sum_{l=0}^{L-2} 
\left(
(\hat c_{\downarrow,l}^\dag \hat c_{\uparrow,l+1}
-\hat c_{\uparrow,l}^\dag \hat c_{\downarrow,l+1}) + \mathrm{h.c.}\right)
.
\end{eqnarray}
Here, $\hat c_{\sigma,l}$ annihilates a fermion with spin $\sigma(=\uparrow, \downarrow)$ at site
$l (=0, 1, \ldots, L-1)$,
$\hat n_{\sigma,l} \equiv \hat c^\dag_{\sigma,l} c_{\sigma,l}$,
$t$ is the nearest-neighbor hopping,
$U$ is the on-site interaction,
$\Delta$ is the coupling to the bulk superconductor,
$\alpha$ is the Rashba-type SOI,
$V_z$ is the Zeeman energy,
and $\mu$ is the chemical potential.
In the following we set $L=200$ and $t=1$.
We take $(\Delta, \alpha, V_z) = (0.1, 0.3, 0.3)$
unless noted otherwise, as in Figs. 2, 3 and 6 of Ref.~\onlinecite{Stoudenmire2011}.
$\epsilon_{\sigma,l} = \epsilon_l$ is the site energy for spin $\sigma$ on site $l$.

For an infinite size system with
$U=\Delta=0$ and $\epsilon_{\sigma, l} = 0$,
it is straightforward to obtain the single-particle dispersion relation
as a function of the quasimomentum $k$:
\begin{equation}
E^{\pm}(k) = t(1-\cos(k)) \pm \sqrt{\alpha^2\sin^2(k) + V_z^2}.
\label{eqn:RZbands}
\end{equation}
In this Rapid Communication we call them the upper and lower Rashba--Zeeman (RZ) bands.

If the Hamiltonian can be mapped to that of a spinless system,
the TS state is realized by the introduction of the pairing $\Delta$.
When $\Delta \ll V_z$
such mapping is possible if
$\mu$ is in only one of the RZ bands. \cite{Lutchyn2010, Oreg2010, Stoudenmire2011}
For more general discussion on the origin of the topological states we refer to Ref.~\onlinecite{Sato2010}.
Note that even when $U=0$, while only quadratic terms of annihilation and creation operators
appear in the Hamiltonian, $\Delta \neq 0$ introduces non-zero matrix elements between
states whose number of fermions differs by two.
The dimension of the Hilbert space
grows exponentially as the number of lattice sites $L$ is increased,
strongly limiting the availability of the exact diagonalization approach.

Suppose that we have a lattice system having two Majorana end modes
$\hat \gamma_1, \hat \gamma_2$
such that $\eta^\pm \equiv \hat \gamma_1 \pm i\hat \gamma_2$ is a fermionic operator
satisfying $(\eta^+)^\dag = (\hat \gamma_1 + i\hat \gamma_2)^\dag
= \hat \gamma_1 - i\hat \gamma_2 = \eta^-$.
If the Majorana operators can be approximated by linear combinations of
the single-particle annihilation and creation operators, 
$\hat \gamma_j = \sum_{\sigma, l}
( a_{\sigma,l}^{(j)}\hat c_{\sigma,l} +  (a_{\sigma,l}^{(j)})^*\hat c^\dag_{\sigma,l})$ ($j=1,2$),
we can think of two single-particle wavefunctions
$|\gamma_j\rangle \equiv \sum_{\sigma,l}(a_{\sigma,l}^{(j)})^*\hat c^\dag_{\sigma,l}|\rangle$,
in which $|\rangle$ is the empty state, as the ``Majorana wavefunctions''.
For the ground state many-body wave functions in the sectors of total number of fermions
being even (e) and odd (o), $|\Psi_\mathrm{e,o}\rangle$ with energies
$E_\mathrm{e,o}$, we can obtain the values of $a_{\sigma,l}^{(j=1,2)}$: \cite{Stoudenmire2011}
\begin{eqnarray}
a_{\sigma,l}^{(1)} &=&
\langle \Psi_\mathrm{o}|\hat c_{\sigma,l}^\dag|\Psi_\mathrm{e} \rangle +
\langle \Psi_\mathrm{e}|\hat c_{\sigma,l}^\dag|\Psi_\mathrm{o} \rangle,\\
a_{\sigma,l}^{(2)} &=&
\langle \Psi_\mathrm{o}|\hat c_{\sigma,l}^\dag|\Psi_\mathrm{e} \rangle -
\langle \Psi_\mathrm{e}|\hat c_{\sigma,l}^\dag|\Psi_\mathrm{o} \rangle.
\end{eqnarray}

In \onlinecite{Stoudenmire2011} the phase diagram for $U>0$ and $U=0$ are obtained by
calculating $|\Psi_\mathrm{e,o}\rangle$ and checking if the following three conditions are met:
(i) $\Delta E\equiv E_\mathrm{e} - E_\mathrm{o}$ vanishes.
(ii) The left reduced density matrices of the system,
obtained from the density matrices $|\Psi_\mathrm{e(o)}\rangle\langle\Psi_\mathrm{e(o)}|$
by tracing out all sites in the right half, have degenerate eigenvalue spectrum.
(iii) $\{a_{\sigma,l}^{(1)}\}$ and $\{a_{\sigma,l}^{(2)}\}$ are spatially
localized to the different ends.
Now we apply these conditions to the case with spatial inhomogeneity.

\begin{figure}[htbp]
\includegraphics{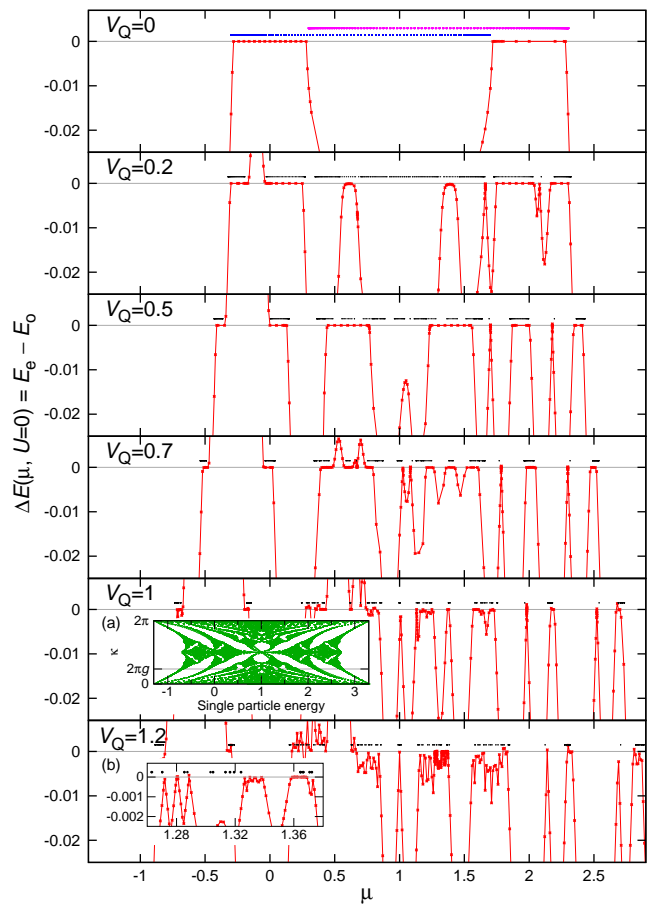}
\caption{(Color online)
Differences in energy between the ground states in
even and odd number sectors of the Hamiltonian plotted against the
chemical potential $\mu$. The energy spectrum of
upper and lower RZ band eigenstates for the $L=233$ case
has been plotted in two lines above the horizontal $\Delta E = 0$ line for $V_\mathrm{Q}=0$,
and the single particle energy spectrum of the $\Delta=0$ case
for $L=200$ has been plotted above the $\Delta E = 0$ line
for each value of $V_\mathrm{Q}=0, 0.2, 0.5, 0.7, 1$ and $1.2$.
The other parameters are $(\Delta, \alpha, V_z) = (0.1, 0.3, 0.3)$ and $U=0$.
Insets: (a) The single particle energy spectrum of the $\Delta=0$ case plotted for
$\kappa \in [0,2\pi]$ and $V_\mathrm{Q}=1$,
and (b) a closeup of the main plot for $\mu \in [1.26:1.38]$ for $V_\mathrm{Q}=1.2$.}
\label{fig:eigv_ediff}
\end{figure}

\noindent\textit{Quasiperiodic site potential.---}
We study the effect of a quasiperiodic site potential, which is given by
\begin{equation}
\epsilon_{\sigma,l} = V_\mathrm{Q} \cos( \kappa (l-l_\textrm{c}) + \delta),
\label{eqn:sitePotential}
\end{equation}
in which $V_\mathrm{Q}\geq 0$, $l_\textrm{c} \equiv (L-1)/2$ and the phase is $\delta = 0$ unless noted.
We choose $\kappa = 2 \pi g$, in which $g = \sqrt{5}-2$.

In the non-interacting case ($U = \Delta = 0$),
we can consider a periodic lattice with $g_n = F_{n-3} / F_{n}$, in which
$F_{n}$ is the $n$-th Fibonacci number, to obtain the approximate eigenenergy distribution.
The Fourier transform of the site potential then has only components with
$k=\pm 2\pi g_n, \pm 2\pi(1-g_n)$ but mixes states between the upper and
lower RZ bands because the spin composition of the states in RZ bands depends
on the wave number.
$g_n$ rapidly converges to $g$ as $n$ is increased, so does the eigenvalue distribution.
As $V_\mathrm{Q}$ is increased, the number and widths of the gaps in the
eigenvalue distribution both increase.
For $V_\mathrm{Q} \sim t$ the spectrum exhibits a distinct self-similar structure
when $\kappa$ is changed, which resembles two Hofstadter butterflies \cite{Hofstadter1976, Kohmoto1983}
shifted in energy and overlapping each other, as shown in the inset (a) of Fig.~\ref{fig:eigv_ediff}.
We note that, for $\alpha = V_z = 0$, all single body wavefunctions are extended for
$V_\mathrm{Q}<t$ and localized for $V_\mathrm{Q}>t$ in the limit of large system
($L\rightarrow \infty$). \cite{Kohmoto1983}

In Fig.~\ref{fig:eigv_ediff} we plot $\Delta E$ against $\mu$ for several
values of $V_\mathrm{Q}$ for $L=200$ sites,
as well as the eigenenergies of the non-interacting case
with the quasiperiodic potential.
The slope of the $\Delta E$ as a function of $\mu$
is often close to $\pm1$ outside the TS phase,
which corresponds to the fact that $N_\mathrm{e} - N_\mathrm{o}$ is close to $\mp1$
in such regions of $\mu$.
The two regions with $\Delta E = 0$ for $V_\mathrm{Q} = 0$ correspond to the ranges of
$\mu$ crossing only one of the two RZ bands.
The RZ bands are gradually mixed and split into several mini-bands as $V_\mathrm{Q}$ is increased
also for the open boundary condition.
While the locations and widths of
some of the regions with vanishing $\Delta E$ resemble those of the mini-bands,
not all the regions of the chemical potential overlapping with one of the mini-bands
have $\Delta E=0$.
While the value of $\Delta E$ outside the plateaus at $\Delta E$
depends on
the choice of $\delta$, the locations and widths of
the plateaus almost do not change when $\delta$ is changed.

\begin{figure}[htbp]
\includegraphics{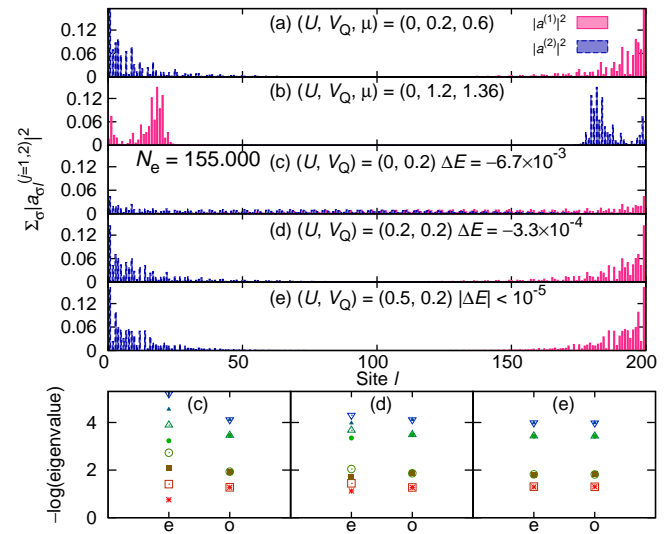}\\
\caption{(Color online)
Top: The values of $\sum_\sigma|a_{\sigma,l}|^2$ are plotted against the site $l$
for
(a) $(U, V_\mathrm{Q}, \mu) = (0, 0.2, 0.6)$,
(b) $(U, V_\mathrm{Q}, \mu) = (0, 1.2, 1.36)$,
(c) $(U, V_\mathrm{Q}, \mu) = (0, 0.2, 0.67448)$,
(d) $(U, V_\mathrm{Q}, \mu) = (0.2, 0.2, 0.75743)$, and
(e) $(U, V_\mathrm{Q}, \mu) = (0.5, 0.2, 0.85472)$.
In (c)--(e), the value of $\mu$ has been chosen so that $N_\mathrm{e} = 155.000$.
Bottom: The eigenvalue spectra of the reduced density matrix for the left half of the
system in states $\Psi_\mathrm{e}$ and $\Psi_\mathrm{o}$ are plotted for
the parameter sets used in (c)--(e).
}
\label{fig:MajoranaModes}
\end{figure}

We have observed that the other two conditions of Ref.~\onlinecite{Stoudenmire2011}, namely
(ii) the localized end states and (iii) the degeneracy of reduced density
matrix eigenvalues, are satisfied in the regions in which $\Delta E$ vanishes,
and they are not when $|\Delta E|\gtrsim 10^{-4}$.
The amplitude distribution of the wave function of the end Majorana modes
for $(V_\mathrm{Q}, \mu) = (0.2, 0.5)$ 
is shown in Fig.~\ref{fig:MajoranaModes}(a).
The distributions of $|a^{(1)}|^2$ and $|a^{(2)}|^2$ are respectively
localized to the right and left ends of the system.
While for $V_\mathrm{Q}\gtrsim 1$, the plots of $\Delta E(\mu, U=0)$ exhibit
significantly shorter plateaus at $\Delta E = 0$,
the two Majorana modes localized at the two ends are observed inside such
plateaus as shown for $(V_\mathrm{Q}, \mu) = (1.2, 1.36)$ in Fig.~\ref{fig:MajoranaModes}(b).
This is a nontrivial observation because for $|V_\mathrm{Q}| > 1$ we expect
localized single-particle states for $\Delta=0$, and localized states would not
support global TSs.
The reduced density matrix eigenvalues are also degenerate, therefore we conclude
that a TS with end Majorana fermions is realized in the $\Delta E=0$ plateaus.
We have also observed that the dependence of the regions with Majorana
end states on the initial phase $\delta$ is weak in our system,
which is similar to the non-interacting case, \cite{Lang2011}
while the sign of $\Delta E$ between these regions depends on the choice of $\delta$.

While we easily obtain the Bogoliubov quasiparticle energies by
retaining the value of $\Delta$ and diagonalizing the Hamiltonian
in the Nambu spinor space, \cite{Oreg2010} the correspondence between
the degenerate region and the single band region in the quasiparticle
spectrum has been observed to be
comparable at best to the correspondence between the former and
the single band region of the mini-bands without $\Delta$.
We believe that not only the energy spectrum but also the spatial
distribution of the states is important in the realization of
the TS states.

\begin{figure}[htbp]
\includegraphics[width=8.66cm]{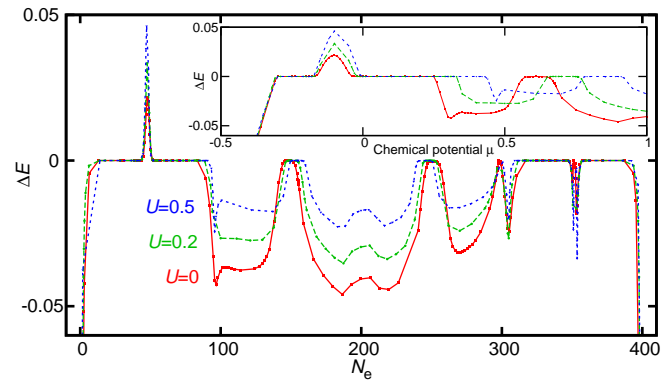}
\caption{(Color online)
The energy difference between ground states in the sectors
with even and odd number of fermions plotted against the number of fermions
for $U=0, 0.2$, and $0.5$.
The other parameters are $(\Delta, \alpha, V_z)=(0.1, 0.3, 0.3)$
and $V_\mathrm{Q} = 0.2$.
Inset: the same plotted against the chemical potential $\mu$.}
\label{fig:Ediff_N_s0.2}
\end{figure}

In the studies of 1D topological phases of \textit{free} fermions
in bichromatic superlattices \cite{Lang2011, Mei2012}
the topological phases appear only at special filling factors
between the bulk bands.
Here we emphasize that in our system of the 1D quantum wire with strong SOI,
placed under a magnetic field and having a proximity-induced pairing,
the topological phases are realized in several finite ranges of
$\mu$ that corresponds to finite ranges of the filling factor,
as observed in Figs.~\ref{fig:eigv_ediff} and \ref{fig:Ediff_N_s0.2} ($U=0$),
some of them much closer to half filling compared to the $V_\mathrm{Q}=0$ case.
This is the main result of this Rapid Communication.

\noindent\textit{Effect of fermion-fermion interaction.---}
Next we study the effect of an on-site fermion-fermion interaction
coexisting with the quasiperiodic site potential modulation.
In Fig.~\ref{fig:Ediff_N_s0.2} the values of $\Delta E(\mu, U)$ for
$U=0, 0.2$ and $0.5$ are plotted against the number of fermions
$N_\mathrm{e}$ in the system as well as against $\mu$.

When $U$ is larger, the amount of increase in $\mu$ required
for adding the same number of fermions at the same filling factor
becomes larger because of the stronger repulsive interaction.
The $\Delta E(\mu)=0$ plateaus in the inset of 
Fig.~\ref{fig:Ediff_N_s0.2} are observed in broader
ranges of the chemical potential, as expected from Ref.~\onlinecite{Stoudenmire2011}
with $V_\mathrm{Q}=0$.
Furthermore, we observe that the plateaus are broader for larger $U$
also in terms of the number of fermions \cite{NumberOfFermions}.

For a fixed $N_\mathrm{e}$, chosen so that
$\Delta E$ approaches zero as $U$ is increased, in Fig.~\ref{fig:MajoranaModes}
(c)--(e)
we observe that the distributions of $|a_{\sigma,l}^{(j)}|$ become
localized toward the ends.
We also observe that the difference between the eigenstate spectra
of the reduced density matrices for $\Psi_\mathrm{e}$ and $\Psi_\mathrm{o}$,
plotted in the bottom part of Fig.~\ref{fig:MajoranaModes},
becomes smaller as $\Delta E$ becomes smaller.
Similar behavior of the distributions of $|a_{\sigma,l}^{(j)}|$ and
the eigenstate spectra of the reduced density matrices are
observed in other plateaus of $\Delta E(\mu, U)$.

\begin{figure}[t]
\includegraphics{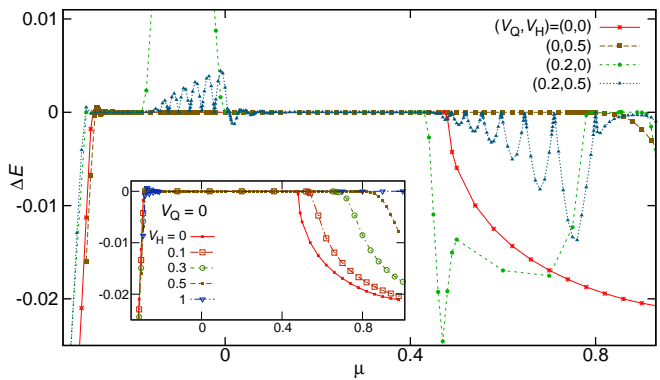}
\caption{(Color online)
$\Delta E$ plotted against $\mu$ for $(V_\mathrm{Q}, V_\mathrm{H})=
(0,0), (0,0.5), (0.2, 0)$, and $(0.2, 0.5)$.
The other parameters are $(\Delta, \alpha, V_z, U)=(0.1, 0.3, 0.3, 0.5)$.
Inset: $\Delta E$ plotted against $\mu$ for $V_\mathrm{Q}=0$ and
$V_\mathrm{H} = 0, 0.1, 0.3, 0.5$ and $1$.}
\label{fig:hrm}
\end{figure}

\noindent\textit{Effect of a trapping potential.---}
In many cold atom experiments, atom clouds are trapped in the vacuum
by (magneto)optical potentials that are better approximated
by a harmonic or Gaussian potential rather than a flat, box-like potential.
Also in condensed matter systems, the shape of the potential for
electrons in the quantum wire would depend on how it is fabricated.
Studying the effect of an additional trapping potential
on the realization of the Majorana end fermions is therefore important and
has already been conducted for non-interacting cases. \cite{Lang2011, Mei2012}
To complete this Rapid Communication, we study the effect of a harmonic potential,
$V_\mathrm{H} [(l-l_\textrm{c})/l_\textrm{c}]^2$ ($l=0,1,\ldots, L-1 = 2l_\textrm{c}$)
added to the site potential (\ref{eqn:sitePotential}).

Figure~\ref{fig:hrm} shows $\Delta E$ plotted against $\mu$ for
$(V_\mathrm{Q}, V_\mathrm{H})=(0,0),(0,0.5),(0.2, 0)$, and $(0.2,0.5)$
with $(\Delta, \alpha, V_z, U)=(0.1, 0.3, 0.3, 0.5)$.
The first three plateaus at $\Delta E = 0$ for $(V_\mathrm{Q}, V_\mathrm{H})=(0.2, 0)$
are in the plot.
We observe that between these plateaus $\Delta E$ plotted for
$(V_\mathrm{Q}, V_\mathrm{H})=(0.2, 0.5)$ shows repeated oscillations.

In the inset of Fig.~\ref{fig:hrm} we have plotted $\Delta E$
for $V_\mathrm{Q}=0$ and several values of $V_\mathrm{H}$.
For increasing $V_\mathrm{H}$ the first region of vanishing $\Delta E$ is broadened.
The Majorana fermion states are localized close to the system
boundary, where the harmonic trap decreases the effective chemical
potential, measured from the local site potential, by almost the
depth of the trap potential $V_\mathrm{H}$.
This explains the increase of the value of $\mu$ at the upper boundary.

On the other hand, our system with the quasiperiodic potential
has the Majorana state broken down not only between the plateau
for the system with $(V_\mathrm{Q}, V_\mathrm{H})=(0.2, 0)$,
but also inside the second and third plateaus, when $V_\mathrm{H}$ is introduced.
We believe that this is because the values of the effective chemical potential
at which the quasiperiodic potential destroys the TS
is reached at some location even for greater values of $\mu$
for $V_\mathrm{H}>0$.
This effect seems to compete with the decrease of the effective chemical potential
at the edge of the fermion distribution and $\Delta E$ oscillates.
While for simplicity we have shown here the results for the case with $U=0.5$,
the discussion above is also consistent with those for other values of $U\geq0$.

In conclusion, we have studied the effect of spatial inhomogeneity
realized by a quasiperiodic site potential modulation applied on a tight-binding 
model of a TS, which is a 1D conductor with SOI
in the proximity of a bulk superconductor and under a magnetic field.
When the modulation is induced, the topological phase appears not
only when the band is almost empty or almost full, but also in
several regions of the filling factor (or the chemical potential) with finite widths
much closer to half filling, even when the modulation is strong enough
to turn the system without SOI and pairing insulating.
We have also studied the effects of on-site fermion-fermion interaction and a harmonic trap.
Without the trap, as the interaction becomes stronger the TS
phase becomes wider in the phase diagram, as in the case without
the spatial inhomogeneity, \cite{Stoudenmire2011} in terms of both
the chemical potential and the number of fermions in the system.
Our results reveal the possibility of realizing
Majorana end states in 1D systems such as cold-atom systems and
superconductor-metal heterostructures
with inherent or imposed inhomogeneties.

This work was partially
supported by the Grant-in-Aid for the Global COE
Program ``The Next Generation of Physics, Spun from
Universality and Emergence'' from MEXT of Japan.
N.~K. is supported by KAKENHI (Grants No. 21540359 and No. 20102008) and JSPS
through its FIRST Program.
Part of the computation in this work has been performed using the facilities of the Supercomputer Center,
Institute for Solid State Physics, University of Tokyo.

Recently, we became aware that
the interplay of disorder and correlation in 1D
TSs has also been investigated in Ref.~\onlinecite{Lobos2012}.


\begin{thebibliography}{99}
\bibitem[]{Read2000} N.~Read and D.~Green, Phys. Rev. B \textbf{61}, 10267 (2000).
\bibitem[]{Kitaev2001} A.~Yu. Kitaev, Phys.-Usp. \textbf{44}, 131, 268 (2001).
\bibitem[]{Ivanov2001} D.~A. Ivanov, Phys. Rev. Lett. \textbf{86}, 268 (2001).
\bibitem[]{Fu2008} L. Fu and C.~L. Kane, Phys. Rev. Lett. \textbf{100}, 096407 (2008).
\bibitem[]{Sato2009} M.~Sato, Y. Takahashi, and S. Fujimoto, Phys. Rev. Lett. \textbf{103}, 020401 (2009).
\bibitem[]{Linder2010}
Y. Tanaka, T. Yokoyama, and N. Nagaosa, Phys. Rev. Lett. \textbf{103}, 107002 (2009);
J. Linder, Y. Tanaka, T. Yokoyama, A. Sudb\o, and N. Nagaosa, Phys. Rev. Lett. \textbf{104}, 067001 (2010);
Y. Tanaka, M. Sato, and N. Nagaosa, J. Phys. Soc. Jpn. \textbf{81}, 011013 (2012);
A. Yamakage, Y. Tanaka, and N. Nagaosa, Phys. Rev. Lett. \textbf{108}, 087003 (2012).
\bibitem[]{Tewari2007} S. Tewari, S. Das Sarma, C. Nayak, C. Zhang, and P. Zoller, Phys. Rev. Lett. \textbf{98}, 010506 (2007).
\bibitem[]{Sau2010} J.~D. Sau, R.~M. Lutchyn, S. Tewari, and S.~Das Sarma, Phys. Rev. Lett. \textbf{104}, 040502 (2010).
\bibitem[]{Alicea2010} J. Alicea, Phys. Rev. B \textbf{81}, 125318 (2010).
\bibitem[]{Lutchyn2010} R.~M. Lutchyn, J.~D. Sau, and S.~Das Sarma, Phys. Rev. Lett. \textbf{105}, 077001 (2010).
\bibitem[]{Oreg2010} Y. Oreg, G. Refael, and F. von Oppen, Phys. Rev. Lett. \textbf{105}, 177002 (2010).
\bibitem[]{Motrunich2001} O. Motrunich, K. Damle, and D.~A. Huse, Phys. Rev. B \textbf{63}, 224204 (2001).
\bibitem[]{Gruzberg2005} I.~A. Gruzberg, N. Read, and S. Vishveshwara, Phys. Rev. B \textbf{71}, 245124 (2005).
\bibitem[]{Brouwer2011} P.~W. Brouwer, M. Duckheim, A. Romito, and F. von Oppen, Phys. Rev. Lett. \textbf{107}, 196804 (2011).
\bibitem[]{Lang2011} L.~-J. Lang, X. Cai, and S. Chen, arXiv:1110.6120.
\bibitem[]{Mei2012} F. Mei, S.~-L. Zhu, Z.~-M. Zhang, C. H. Oh, and N. Goldman, Phys. Rev. A \textbf{85}, 013638 (2012).
\bibitem[]{Braunecker2010} B. Braunecker, G.~I. Japaridze, J.~Klinovaja, and D.~Loss, Phys. Rev. B \textbf{82}, 045127 (2010).
\bibitem[]{Fidkowski2010} L. Fidkowski and A. Kitaev, Phys. Rev. B \textbf{81}, 134509 (2010).
\bibitem[]{Gangadharaiah2011} S. Gangadharaiah, B. Braunecker, P. Simon, and D. Loss, Phys. Rev. Lett. \textbf{107}, 036801 (2011).
\bibitem[]{Stoudenmire2011} E.~M. Stoudenmire, J. Alicea, O.~A. Starykh, and M.~P.~A. Fisher, Phys. Rev. B \textbf{84}, 014503 (2011).
\bibitem[]{Sela2011} E. Sela, A. Altland, and A. Rosch, Phys. Rev. B \textbf{84}, 085114 (2011).
\bibitem[]{Lutchyn2011b} R.~M. Lutchyn and M.~P.~A. Fisher, Phys. Rev. B \textbf{84}, 214528 (2011).
\bibitem[]{Roati2008} G. Roati \textit{et al.}, Nature \textbf{453}, 895 (2008).
\bibitem[]{Lucioni2011} E. Lucioni \textit{et al.}, Phys. Rev. Lett. \textbf{106}, 230403 (2011).
\bibitem[]{Potter2010} A.~C. Potter and P.~A. Lee, Phys. Rev. Lett. \textbf{105}, 227003 (2010); Phys. Rev. B \textbf{83}, 094525 (2011).
\bibitem[]{Lutchyn2011a} R.~M. Lutchyn, T.~D. Stanescu, and S.~Das Sarma, Phys. Rev. Lett. \textbf{106}, 127001 (2011);
T.~D. Stanescu, R.~M. Lutchyn, and S.~Das Sarma, Phys. Rev. B \textbf{84}, 144522 (2011).
\bibitem[]{White1992} S.~R. White, Phys. Rev. Lett. \textbf{69}, 2863 (1992); Phys. Rev. B \textbf{48}, 10345 (1993).
\bibitem[]{Schollwock2011} U.~Schollw\"ock, Ann. Phys. \textbf{326}, 96 (2011).
\bibitem[]{Sato2010} M. Sato, Y. Takahashi, and S. Fujimoto, Phys. Rev. B \textbf{82}, 134521 (2010).
\bibitem[]{Kohmoto1983} M.~Kohmoto, Phys. Rev. Lett. \textbf{51}, 1198 (1983); C. Tang and M. Kohmoto, Phys. Rev. B \textbf{34}, 2041 (1986).
\bibitem[]{Hofstadter1976} D.~R. Hofstadter, Phys. Rev. B \textbf{14}, 2239 (1976).
\bibitem[]{NumberOfFermions} $N_\mathrm{e}$ and $N_\mathrm{o}$ are both increasing functions of $\mu$.
While they are degenerate only inside the topological insulator phase,
$|N_\mathrm{e}-N_\mathrm{o}|$ is observed to be at most unity.
Therefore the widths of the $\Delta E=0$ plateaus are not
significantly changed if we choose to plot $\Delta E$ against $N_\mathrm{o}$.
\bibitem[]{Lobos2012} A.~M. Lobos, R.~M. Lutchyn, and S. Das Sarma, arXiv:1202.2837.
\end{thebibliography}
\end{document}